\documentclass[prl,aps,showpacs,twocolumn,floatfix]{revtex4-1}
\usepackage{graphicx}
\usepackage{bm}
\usepackage{epsfig}
\usepackage{amsmath}
\usepackage{amssymb}

\newcommand{\eps}{\varepsilon}
\newcommand{\al}{\alpha}

\newcommand{\gm}{\gamma}

\newcommand{\sg}{\sigma}

\newcommand{\lm}{\lambda}

\newcommand{\eq}{\begin{eqnarray}}
\newcommand{\eqx}{\end{eqnarray}}
\newcommand{\bal}{\begin{align}}
\newcommand{\eal}{\end{align}}
\newcommand{\ba}{\begin{equation}}
\newcommand{\ea}{\end{equation}}
\newcommand{\fr}{\frac}

\newcommand\n{\nonumber \\}

\begin{document}

\title{A QCD interpretation of the scaling observed in the LHC proton-proton
  elastic cross-sections at moderate transfer momentum}
\author{R. Peschanski}
\email{robi.peschanski@cea.fr}
\author{B. G. Giraud}
\email{bertrand.giraud@cea.fr}
\affiliation{Institut de Physique Th\'eorique,
Centre d'Etudes de Saclay, 91191 Gif-sur-Yvette, France}
\begin{abstract}
A phenomenological scaling property of the LHC proton-proton elastic
scattering cross sections at moderate transfer momentum has been recently
observed. The theoretical QCD saturation framework for hard elastic scattering
on the proton, which we recall, predicts a strikingly similar scaling property
in the same configuration of energy and momentum transfer. This similarity
favors a theoretical interpretation where the hard amplitude is the building
block of the $pp$ amplitude via unitarization. The values of the scaling
exponents favor two QCD saturation approaches where the saturation scale
exponent is of order one half of that which is found for hard elastic
scattering and deep inelastic processes on the proton.
\end{abstract}

\date{\today}
\maketitle

Recently \cite{scaling}, scaling properties of proton-proton 
elastic scattering at LHC energies have been found, based on 
the elastic differential cross-sections measured by the TOTEM collaboration
\cite{TOTEM}. Conveniently scaled with energy, the cross sections
appear to be  dependent upon one variable only, within error bars. One finds
\ba
{\fr {d\sg^{pp}}{dt}}(s,t) \sim  s^{\alpha}  \ 
f_{pp}\left[|t|^{\gm} s^{\eps}\right]\ ,
\label{crossscalingzero}
\ea
with $\al=0.305$, $\gm=0.72$ and $\eps=0.065$. The usual Mandelstam
variables are $s$,  the center-of-mass (c.o.m.) energy squared,
and $t$, minus the square of the  momentum transfer $q$.  The
parameterization of the corresponding amplitude, proposed in
\cite{scaling}, uses a scaling variable having the dimension of a transverse
momentum squared, namely
\ba
\tau\ = \ |t|\times (s/ { {\rm TeV^2}})^{\lm} \ ,
\label{scalingtau}
\ea
where $\lm = \eps/\gm = 0.09$.
The scaling property is observed in the LHC range for c.o.m. energies
$\sqrt{s} = 2.76,7,8$ and $13\ {\rm TeV}$ in a moderate  range of momentum
transfer $|t|\sim  [.3,.9]\ {\rm GeV}^2$, spanning the so-called dip-bump
region of the $pp$ cross-sections.

Let us consider another high energy elastic scattering case.  This one can
be computed in the resummed perturbative QCD framework at weak coupling. It
describes the scattering of a hard probe of typical momentum $k$ on the
proton  of typical soft momentum $k_0$ at {\it moderate} momentum transfer
$q=\sqrt{|t|}$, in the range
\ba
\sqrt{s} >> k >> q >> k_0\ .
\label{intermediate}
\ea
The QCD calculation of the corresponding amplitude has been performed
\cite{Marquet} in the framework of the Balitsky-Kovchegov (BK) equation
\cite{BK} where the linear non-forward BFKL evolution equation
\cite{Lipatov} is corrected for the nonlinear effect of gluon recombination.
The result for the hard high energy elastic cross-section in the range
\eqref{intermediate} exhibits a typical scaling property.
\ba
{\cal A}(s,s_0,q^2,k^2) \propto  i\ s 
\left( q^2 / k^2\right)^{\gm_c} \left(s / s_0\right)^{\eps_c}   \ ,
\label{BKamplitude}
\ea
where $\sqrt{s_0}$ is a threshold energy. The scaling exponents $\gm_c,\eps_c$
are obtained  \cite{Traveling} from the implicit equations
\ba
{\chi(\gm_c) }= \gm {\fr {d\chi(\gm)}{d\gm}}\Big\vert_{\gm_c}, \quad \eps_c =
\bar \al\ \chi(\gm_c) ,
\label{indices}
\ea
where $\bar \al$ is the (fixed) QCD coupling constant and $\chi(\gm) $ is the
BFKL kernel \cite{Lipatov}.  Some correction factors, including logarithmic
scaling terms, can also be computed  \cite{Marquet} but have been omitted in
Eq.\eqref{BKamplitude}, for simplicity at this stage. The $t$-dependent
coupling to the proton is assumed to be almost constant in the scaling range.
It is quite important to note that the solution given by equations
(\ref{BKamplitude}) and (\ref{indices}) extend beyond BK to any nonlinear
damping factors \cite{munier} and also to modifications of the BFKL kernel
due to renormalization group improvements \cite{enberg}.

The cross-section one obtains then reads
\ba
{\fr {d\sg^{BK}}{dt}}(s,t) \sim  \ 
f_{BK} \left[|t|^{\gm_c} s^{\eps_c}\right]\ .
\label{qcdscaling}
\ea
with a scaling variable of a strikingly similar form to
\eqref{crossscalingzero}, with exponents $\gm_c, \eps_c $ which will be
examined later on by comparison with those  $\gm, \eps $ of formula
\eqref{crossscalingzero}. The exponent $\al$ of the normalization prefactor
will be discussed later.

The main lesson of the analysis \cite{Marquet} is that the saturation scale
squared,
\ba
Q_s^2(s) =  |t| \times \left(s / s_0\right)^{\lm_s} , \quad  \lm_s = \eps_c/\gm_c
\ ,
\label{saturationscale}
\ea 
governs the scaling properties of the amplitude \eqref{BKamplitude}.
It has the physical meaning of the momentum scale (e.g. inverse size)  of the
color coherence domain in the proton for the scattering at fixed momentum
transfer $q$. By contrast with the forward case, where a phenomenological 
input $Q_0$ has to be introduced, the scale is fully defined by the scattering
kinematics. However, the saturation exponent $\lm_s$ is expected to remain
the same. 

The scaling properties of the cross-sections \eqref{crossscalingzero},
\eqref{qcdscaling} and of the characteristic energy dependence of the transfer
momentum scale \eqref{scalingtau}, \eqref {saturationscale} are very similar,
taking into account that the kinematic range of validity variables is
essentially the same, namely  high c.o.m. energy and  moderate momentum
transfer.  We may then ask whether the hard scattering amplitude
can be related to the LHC $pp$ elastic scattering cross-section.

Obviously the amplitudes themselves are very different. The $pp$ cross-section
has a decreasing exponential behaviour. Indeed, the parameterization of the
scaling elastic scattering amplitude fitted in \cite{scaling} is a combination
of two decreasing exponentials of the scaling variable $\tau$. The BK 
amplitude Eq.\eqref{BKamplitude} is an increasing  power-like function of the
saturation scale $Q_s$.  We infer that if the amplitude \eqref{BKamplitude} is
the building block of the $pp$ elastic amplitude, e.g. in an iterative
unitarization procedure, it will preserve the scaling property. This was
indeed observed in \cite{romuald}, where elastic scattering was expressed
through the AdS/CFT correspondence, using a scaling correlator of Wilson loops
as a phase shift being the building block of the amplitude in impact parameter
space, through eikonal exponentiation. However, in our case, unitarization
is expected to be local in transverse momentum and not in impact parameter,
in order to preserve the observed scaling properties in the appropriate
moderate $t$ range. The exponentiation in transfer momentum variable is at the
root of the phenomenological description of data \cite{scaling}. We leave the
theoretical study of such a unitarization procedure for further study.

Looking for a QCD interpretation of the scaling in $pp$, based upon an
amplitude of the form \eqref{BKamplitude}, we now examine the values of the 
scaling exponents $\gm_c,\eps_c$, and of their ratio 
$\eps_c/\gm_c= \lm_s$ governing the saturation scale dependence on energy.
We compare them respectively to $\gm,\ \eps$ and $\lm$, of
Eq.\eqref{crossscalingzero}. For this, we may use the determination of the
scaling exponents obtained from
the description of exclusive vector meson production at HERA as a function of
the transverse momentum \cite{vector}. The amplitude \eqref{BKamplitude} is an
essential input, apart from small fitting adjustments, for this quasi-elastic
process initiated by the hard probe of a virtual photon. One finds
$\gm_c \sim 0.74, \quad \lm_s \sim 0.22, \quad \eps_c=\gm_c \lm_s \sim 0.16\ $.
Note that, as expected from the theory \cite{Marquet}, the value of $\lm_s$
agrees with that of the forward saturation case in deep inelastic scattering
\cite{munier}. Such values deserve some comment.

The value of $\gm_c  \approx 0.74$  is remarkably close to  that,
$\gm \approx 0.72$ in Eq.\eqref{crossscalingzero} but slightly differs from
the theoretical BK value, $\sim 0.63$. The value obtained for $\lm_s$ is in
agreement with many determinations of the saturation scale evolution in deep
inelastic processes \cite{IIM}. The value of $\lm_s \approx 0.22$ is about
twice that of $\lm \approx 0.09$ which one obtains for the observed scaling
in $pp$ elastic collisions, see \eqref{scalingtau}. The question now is how
one may interpret such results, in particular the last one.

We are led to consider theoretical explanations in the QCD interpretation
coming from the saturation approach of hard elastic scattering. The main point
is to explain the factor of about $\frac{1}{2}$ in the parameter $\lm$ with
respect to $\lm_s$. To our knowledge, two theoretical approaches can lead to
this property for the saturation scale.

The first approach comes from renormalization group improvements of the
BK equation \cite{enberg}. Remarkably all considered schemes lead to a quite
unified set of exponents, where $\lm_s$ is around half of the leading logarithm
value, in agreement with the value of $\lm$ in \eqref{crossscalingzero}.
Concerning $\gm_c$, the same approach provides $\gm_c \in [0.7,0.8] $,
compatible with the value seen in \eqref{crossscalingzero}.
Note that, on a phenomenological ground, the new value,
$\eps_c = \lm_c \ \gm_c \approx 0.08$, is rather close to the $\eps$ value
$0.09$. It is also in agreement with the effective energy exponent of the
$pp$ total cross-section at the LHC \cite{exponent}. In this framework,
hard probes of momentum $k$, see \eqref{intermediate}, are present in either
of the two protons, leading to the desired hard elastic scattering.

A second possible approach is based upon a different QCD mechanism \cite{levin},
by a combination with the hard scattering properties derived in
\cite{Marquet}. In this framework, a hard probe of typical momentum $k >> q$
appears at an intermediate value $y$ in the rapidity distance $Y= \ln (s/s_0)$
between the two protons. It is acting with rapidity $y= \ln (s'/s_0)$ on
one proton and with rapidity $Y-y= \ln (s/s')$ on the other one.  The key
property we use \cite{levin} is that the interaction is described by the
product of amplitudes of the hard probe to each of the incident protons at
given transverse momentum. Using the scaling amplitudes of the form
\eqref{BKamplitude} derived in Ref.\cite{Marquet} in this framework leads to
\eq
{\cal A}(s,s',t=-q^2,k^2)\ \times\ {\cal A}(s',s_0,t=-q^2,k^2)\ \propto \n
\left( q^2 / k^2\right)^{2\gm_c} \left(s / s_0\right)^{\eps_c} 
=\{\left( q^2 / k^2\right)^{\gm_c} \left(s / s_0\right)^{\eps_c/2}\}^2 \ .
\label{doubleBKamplitude}
\eqx    
Accordingly, the scaling exponents of \eqref{doubleBKamplitude} boil down to
$\gm = \gm_c, \eps = \eps_c/2$. Then the saturation scale exponent is
$\lm = \eps_c/(2\gm_c)= \lm_s/2.$ The factor $1/2$ was already noticed in
\cite{levin}.  Note that this result is, in the case of the resulting QCD
amplitude of \cite{Marquet}, independent of the position in rapidity of the
hard probe. This rapidity $y$ disappears in formula \eqref{doubleBKamplitude}.
Note that the dependence on $k$ can be factorized out.

The question which now arises is the identification of the hard probes
appearing in $pp$ scattering. Such probes could be identified with the
so-called "hot spots" \cite{Mueller:1990er,Bartels:1991tf,Albacete:2016pmp}.
Hot spots are defined as small regions of high gluon density in an interacting
particle at high energy. They are associated with the mechanism of gluon
branching and recombination which is characteristic of QCD processes at high
energy and moderate but high enough virtuality to justify a resummed
perturbative framework of QCD. Hence we are led to the conjecture that the
identification of the scaling properties of $pp$ elastic scattering with
those in hard elastic scattering in the same kinematic configuration may
provide a new evidence for the existence hot spots. Our quantitative study
and its QCD interpretations confirm the suggestion by the original paper
\cite{scaling}. In this framework, we would interpret the exponent $\al$,
in the scaling equation \eqref{crossscalingzero}, as parameterizing
the energy dependence of the overall number of hot spots.

In conclusion, we found a link between the scaling properties of the measured
LHC $pp$ differential cross sections at moderate momentum transfer and
those of hard elastic scattering in the same kinematics. It leads to two
existing QCD interpretations in the saturation framework. It reinforces the
legitimity of hot spot models.

{\it Acknowledgements} Thanks are due to Christian Baldonegro, Christophe Royon
and Anna Stasto for their communication of the experimental results and
their discovery of scaling in them. Thanks are also due to Christophe Royon
and Romuald Janik for fruitful discussions.


\begin{thebibliography}{99}

\bibitem{scaling}
C.~Baldenegro, C.~Royon and A.~M.~Stasto,
``Scaling properties of elastic proton-proton scattering at LHC 
energies,''
Phys. Lett. B \textbf{830} (2022), 137141
[arXiv:2204.08328 [hep-ph]].

\bibitem{TOTEM}
G.~Antchev \textit{et al.} [TOTEM],
``Measurement of proton-proton elastic scattering and total 
cross-section at S**(1/2) = 7-TeV,''
EPL \textbf{101} (2013) no.2, 21002.
\newline
$ib.$ ``Evidence for non-exponential elastic proton\textendash
{}proton differential cross-section at low |t| and $\sqrt{s}$=8 
TeV by TOTEM,''
Nucl. Phys. B \textbf{899} (2015), 527-546.
\newline
$ib.$ ``Elastic differential cross-section ${\mathrm{d}}\sigma 
/{\mathrm{d}}t$ at $\sqrt{s}=2.76\hbox { TeV}$ and implications on 
the existence of a colourless C-odd three-gluon compound state,''
Eur. Phys. J. C \textbf{80} (2020) no.2, 91.
\newline
$ib.$ ``First determination of the ${\rho }$ parameter at 
${\sqrt{s} = 13}$ TeV: probing the existence of a colourless C-odd 
three-gluon compound state,''
Eur. Phys. J. C \textbf{79} (2019) no.9, 785.

\bibitem{Marquet}
C.~Marquet, R.~B.~Peschanski and G.~Soyez,
``Traveling waves and geometric scaling at non-zero momentum transfer,''
Nucl. Phys. A \textbf {756} (2005) 399-418, 
[arXiv:hep-ph/0502020 [hep-ph]].

\bibitem{BK}
I.~Balitsky,
``Operator expansion for high-energy scattering,''
Nucl. Phys. B \textbf{463} (1996), 99-160
[arXiv:hep-ph/9509348 [hep-ph]].
\newline
Y.~V.~Kovchegov,
``Unitarization of the BFKL pomeron on a nucleus,''
Phys. Rev. D \textbf{61} (2000), 074018
[arXiv:hep-ph/9905214 [hep-ph]].

\bibitem{Lipatov}
L.~N.~Lipatov,
``The Bare Pomeron in Quantum Chromodynamics,''
Sov. Phys. JETP \textbf{63} (1986), 904-912
Leningrad-85-1137.

\bibitem{Traveling}
S.~Munier and R.~B.~Peschanski,
``Geometric scaling as traveling waves,''
Phys. Rev. Lett. \textbf{91} (2003), 232001
[arXiv:hep-ph/0309177 [hep-ph]].

\bibitem{munier}
S.~Munier and R.~B.~Peschanski,
``Traveling wave fronts and the transition to saturation,''
Phys. Rev. D \textbf{69} (2004), 034008
[arXiv:hep-ph/0310357 [hep-ph]].

\bibitem{enberg}
R.~Enberg,
``Traveling waves and the renormalization group improved Balitsky-Kovchegov
equation,''
Phys. Rev. D \textbf{75} (2007), 014012
[arXiv:hep-ph/0612005 [hep-ph]].

\bibitem{romuald}
R.~A.~Janik and R.~B.~Peschanski,
``High-energy scattering and the AdS / CFT correspondence,''
Nucl. Phys. B \textbf{565} (2000), 193-209
[arXiv:hep-th/9907177 [hep-th]].

\bibitem{vector}
C.~Marquet, R.~B.~Peschanski and G.~Soyez,
``Exclusive vector meson production at HERA from QCD with saturation,''
Phys. Rev. D \textbf{76} (2007), 034011
[arXiv:hep-ph/0702171 [hep-ph]].


\bibitem{IIM}
E.~Iancu, K.~Itakura and S.~Munier,
``Saturation and BFKL dynamics in the HERA data at small x,''
Phys. Lett. B \textbf{590} (2004), 199-208
[arXiv:hep-ph/0310338 [hep-ph]].


\bibitem{exponent}
M.~Broilo, D.~A.~Fagundes, E.~G.~S.~Luna and M.~Pel\'aez,
``Soft Pomeron in light of the LHC correlated data,''
Phys. Rev. D \textbf{103} (2021) no.1, 014019
[arXiv:2012.08664 [hep-ph]]

\bibitem{levin}
C.~Contreras, E.~Levin and M.~Sanhueza,
``Soft pomeron in the color glass condensate approach,''
Phys. Rev. D \textbf{106} (2022) no.3, 034011
[arXiv:2203.10296 [hep-ph]].


\bibitem{Mueller:1990er}
A.~H.~Mueller,
``Parton distributions at very small x values,''
Nucl. Phys. B Proc. Suppl. \textbf{18} (1991), 125-132.


\bibitem{Bartels:1991tf}
J.~Bartels, A.~de Roeck and M.~Loewe,
``Measurement of hot spots inside the proton at HERA and LEP/LHC,''
Z. Phys. C \textbf{54} (1992), 635-642.


\bibitem{Albacete:2016pmp}
J.~L.~Albacete and A.~Soto-Ontoso,
``Hot spots and the hollowness of proton\textendash{}proton interactions at
high energies,''
Phys. Lett. B \textbf{770} (2017), 149-153
[arXiv:1605.09176 [hep-ph]].

\end{thebibliography}
\end{document}